\documentclass[prl,final,amsmath,twocolumn,showpacs,superscriptaddress]{revtex4-1}
\usepackage{graphicx}
\usepackage{dsfont}
\usepackage{bm}
\usepackage[colorlinks,linkcolor=blue,citecolor=blue]{hyperref} 
\usepackage{epstopdf}
\usepackage{dcolumn}
\usepackage{epsfig}
\usepackage{subfigure}

\begin{document}

\title{Random matrix theory for underwater sound propagation}

\author{Katherine C.~Hegewisch}
\affiliation{Department of Physics and Astronomy, Washington State University, Pullman, Washington,99164-2814}
\author{Steven Tomsovic}
\altaffiliation[Permanent address:]{Department of Physics and Astronomy, PO Box
642814, Washington State University, Pullman, WA 99164-2814, USA}
\affiliation{Department of Physics, Indian Institute of Technology Madras, 
Chennai, 600036, India}

\date{\today} 

\begin{abstract}

Ocean acoustic propagation can be formulated as a wave guide with a weakly random medium generating multiple scattering.  Twenty years ago, this was recognized as a quantum chaos problem, and yet random matrix theory, one pillar of quantum or wave chaos studies, has never been introduced into the subject.  The modes of the wave guide provide a representation for the propagation, which in the parabolic approximation is unitary.   Scattering induced by the ocean's internal waves leads to a power-law random banded unitary matrix ensemble for long-range deep ocean acoustic propagation.  The ensemble has similarities, but differs, from those introduced for studying the Anderson metal-insulator transition.  The resulting long-range propagation ensemble statistics agree well with those of full wave propagation using the parabolic equation.

\end{abstract}
\maketitle

Underwater sound provides a means of remote sensing the ocean interior and monitoring global climate change~\cite{Flatte79,Munk95} amongst other motivations.  Beginning in the latter part of the 1980's decade it was realized that the ray dynamics underlying ocean acoustic propagation in many contexts was chaotic~\cite{Palmer88,Palmer91,Smith92a,Smith92b} and therefore the field serves as a domain with several unique features for studies of wave/quantum chaos~\cite{Brown03,Beronvera03,Tomsovic10}.  To date, one of the theoretical foundations of quantum chaos, random matrix theory (RMT)~\cite{Brody81,Bohigas84,Mehta04}, has not been applied previously in the ocean acoustic propagation context.  This is in stark contrast to other linear acoustics fields, where random matrix theory has been of growing usefulness since its introduction over twenty years ago~\cite{Weaver89,Wright10}.  RMT has the potential to improve the understanding of many of the observed statistical behaviors and connect them to parameters of the ocean environment, and give a new more efficient method of making simulations.  Conversely, unique features of ocean acoustics have the potential to influence the future development of RMT through the introduction of new models.

The earliest RMT application to elastodynamics~\cite{Weaver89} had a direct experimental connection to the classical, structureless Gaussian/circular ensembles of Wigner and Dyson~\cite{Mehta04}.  Ocean acoustic propagation cannot be represented in this way.  The purpose of this letter therefore is to step through the construction of a structured random matrix ensemble appropriate for ocean acoustic propagation in as straightforward a physical context as possible.  For simplicity, long range propagation at a fixed, low angular frequency $\omega$ is considered.  The sound channel creates a vertical wave guide~\cite{Munk74} in which the ocean's internal waves~\cite{Garrett79} generate multiple scatterings (introducing chaos).  Horizontal out-of-plane scattering can be neglected, as can absorption, surface, and bottom interactions.  Furthermore, the dominant effect of internal waves is small angle, forward scattering.  Incorporation of these approximations into the appropriate scalar wave equation leads to a paraxial optical (parabolic) equation.  It has a direct analogy to the Schr\"odinger equation, making the wave and quantum chaos connection even closer.  In this analogy, the wave vector inverse $k_0^{-1}$ plays the role of a pseudo-Planck's constant and the propagation range $r$ the role of a pseudo-time, i.e.
\begin{equation}
\label{pwe}
 \frac{i}{k_0} \frac{\partial \Psi(z;r)}{\partial r} = \left(
   -  \frac{1}{2k_0^2} \frac{\partial^2}{\partial z^2} + V(z,r) \right) \Psi(z;r) \ ,
\end{equation}
where the real part of $\Psi(z,r)$ multiplied by a traveling phase and $r^{-1/2}$ is the sound pressure amplitude.  The ``potential'' is
\begin{equation}
\label{U}
V(z,r) = \frac{1}{2} \left[ 1 - \frac{c_0^2}{c^2(z,r)} \right] = V_0(z) + \epsilon V_1(z,r)\ ,
\end{equation}
$c_0$ is a reference sound speed $\approx 1.5 $km/s, and $k_0^{-1}=c_0/\omega$.  $V_0(z)$ typically has a minimum at approximately $1$ km depth and is modeled with a Munk profile~\cite{Munk74}.  It possesses an exponential form near the surface due to temperature decrease and a linear dependence deep underwater due to pressure increase.  Multiple scattering is induced by the internal wave fluctuations in the range and depth dependences $V_1(z,r)$ and the form used here is that of Ref.~\cite{Colosi98}.  More details and further references can be found in the review paper~\cite{Brown03}.  Note that our calculations have no cutoffs in mode number for the ocean surface or bottom.

Wave field propagation for a parabolic equation is unitary and the modes provide a physically relevant basis with which to study the effects of scattering~\cite{Dozier78,Dozier78b}.  Extensions to adiabatically defined modes could also be incorporated, but here the modes are independent of range.  The unitary propagation matrix $U$ expressed in the mode basis is the natural vehicle for which to construct the random matrix ensemble.  The problem is to identify and incorporate into the ensemble's structure all information that survives long-range propagation (up to several thousand kilometers) and no more.

Without scattering from internal waves, or indeed any plausible scattering mechanism, the unitary propagation matrix would be a diagonal matrix, $U^{diag}=\Lambda$, and accumulate a phase proportional to the range propagated, 
\begin{equation}
\Lambda_{mn}(r)=\delta_{mn}\exp(-ik_0 E_m r)\ ,
\end{equation}
where $E_m$ is the energy of the $m^{th}$ mode of the vertical waveguide.  To propagate an initial wave field just requires its initial modal decomposition and these phases.  

Perturbatively to first order, internal waves generate a propagator
\begin{equation}
U\approx \Lambda \left( I - i\epsilon k_0\int^r_0 {\rm d}r^\prime V_I \right) \ ,
\end{equation}
which requires restoration of unitarity ($V_I$ is the operator corresponding to $V_1(z,r)$ in the interaction picture).   One technique applies the Cayley transform of a Hermitian operator $A$
\begin{equation}
U=\Lambda (I+i\epsilon A)^{-1}(I-i\epsilon A) \ ,\  A= \frac{k_0}{2} \int^r_0 {\rm d}r^\prime V_I \ ,
\label{ensemble}
\end{equation}
which is quite useful because the internal waves generate a weak scattering locally and evaluating the operator inverse is straightforward.  As a consequence, it is natural to construct a fixed-range building block unitary propagator $U_b=U(r=r_b)$ for ranges long enough for the internal waves to generate sufficiently random behaviors, but short enough that perturbation theory is still a viable approximation.  Transfer matrix building blocks are used in Ref.~\cite{Perez07}.  Long range propagation follows by multiplying the requisite number of independently drawn members of a $U_b$ ensemble to arrive at an ensemble of $U$ for the full range [i.e.~$U(r=\ell r_b)=\prod^\ell_j U_{b,j}$].

\begin{figure}
\centering
     \subfigure{
          \includegraphics[width= 3.5 in,angle=0]{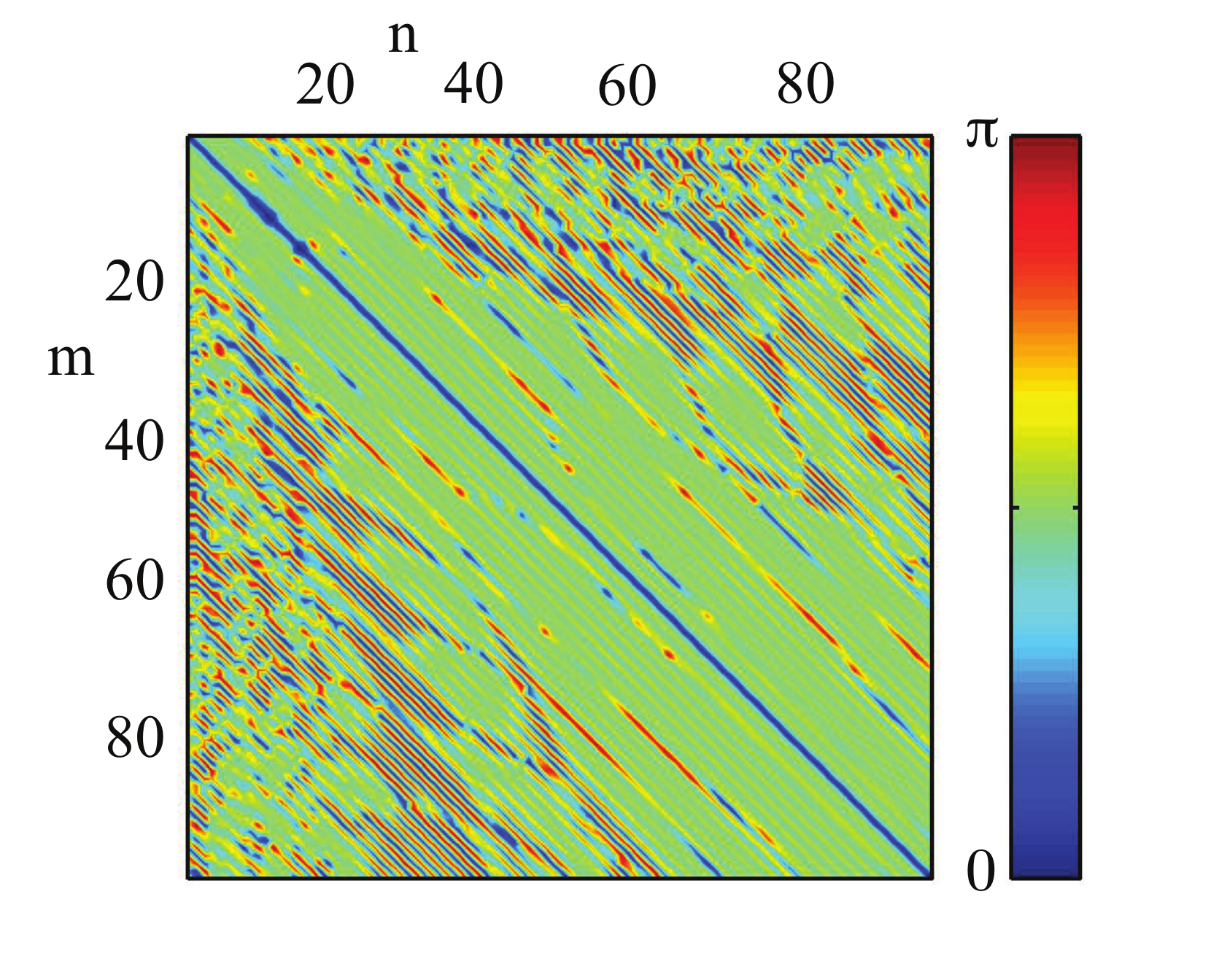}} 
     \subfigure{
          \includegraphics[width= 3.5 in,angle=0]{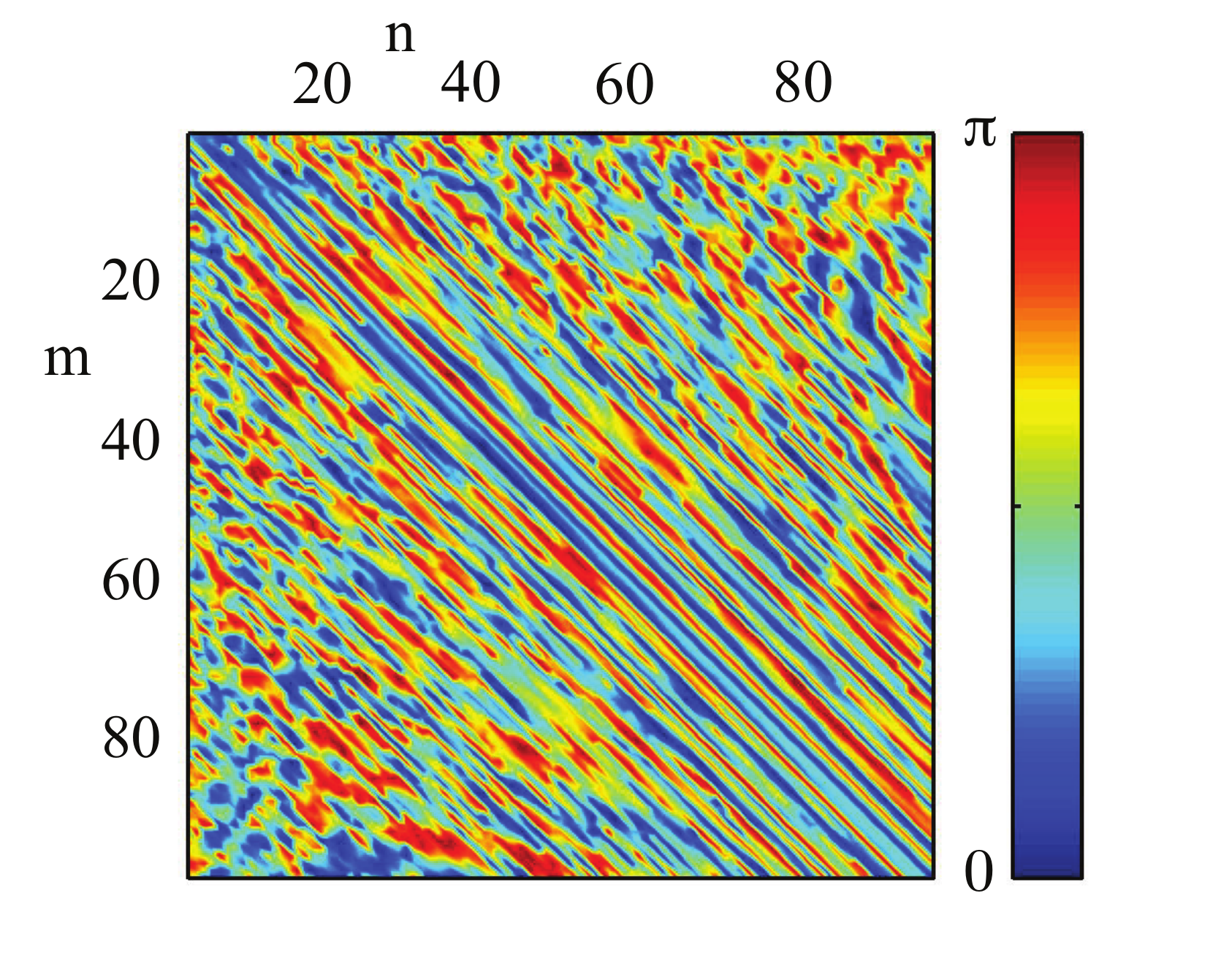}}
\caption{The phases of the matrix elements of $\Lambda^{-1/2} U \Lambda^{-1/2} $ for propagation through a single internal wave field.  The top panel is for $1$ km, and the bottom panel $50$ km.  The nonrandom behaviors disappear with increasing range.}
\label{fig1}
\end{figure}

Hence, ideally there are three statistical properties of interest to determine a practical value for the range $r_b$:  i) correlations between a matrix element of $U_b$ for a given block and its adjacent block should be small, ii) the phases of each $U_{b,mn}$ should be largely randomized, and iii) ideally, dynamical correlations between neighboring matrix elements should be minimal.  For extremely short range propagation, there is little scattering, $U$ is nearly diagonal, and the matrix elements are highly correlated with little randomness.  As the propagation range increases, there comes a point at which the  phases of the $U_b$ matrix elements become more or less randomized. Figure~\ref{fig1} shows
the randomization is well underway by $50$ km at $75$ Hz (used throughout).  At this range the correlations between $U_{b,mn}$ for adjacent blocks have fallen to roughly $10\%$ or less.  Conveniently, this is roughly the range at which wave energy has cycled once from the upper to lower turning point and back again.  As the strongest perturbations are close to the surface, choosing this range is consistent with accounting for one strong perturbation cycle.  Other than effects due to the unitarity constraints, the matrix elements are beginning to behave like zero-centered, complex, Gaussian random variables.  Thus, the $A_{mn}$ are taken as independent, complex, Gaussian random variables; the diagonal elements are real.  The dynamical correlations amongst neighboring matrix elements are not so small, especially along diagonals of $U_b$; `dynamical' means correlations in addition to those induced by unitarity.  Nevertheless, as a starting point, these correlations are ignored.  If later, it is found that the ensemble is in some way deficient, one could revisit their incoporation.  That leaves just the variance determination for each matrix element.

\begin{figure}
\centering
     \subfigure{
          \includegraphics[width= 3.5 in,angle=0]{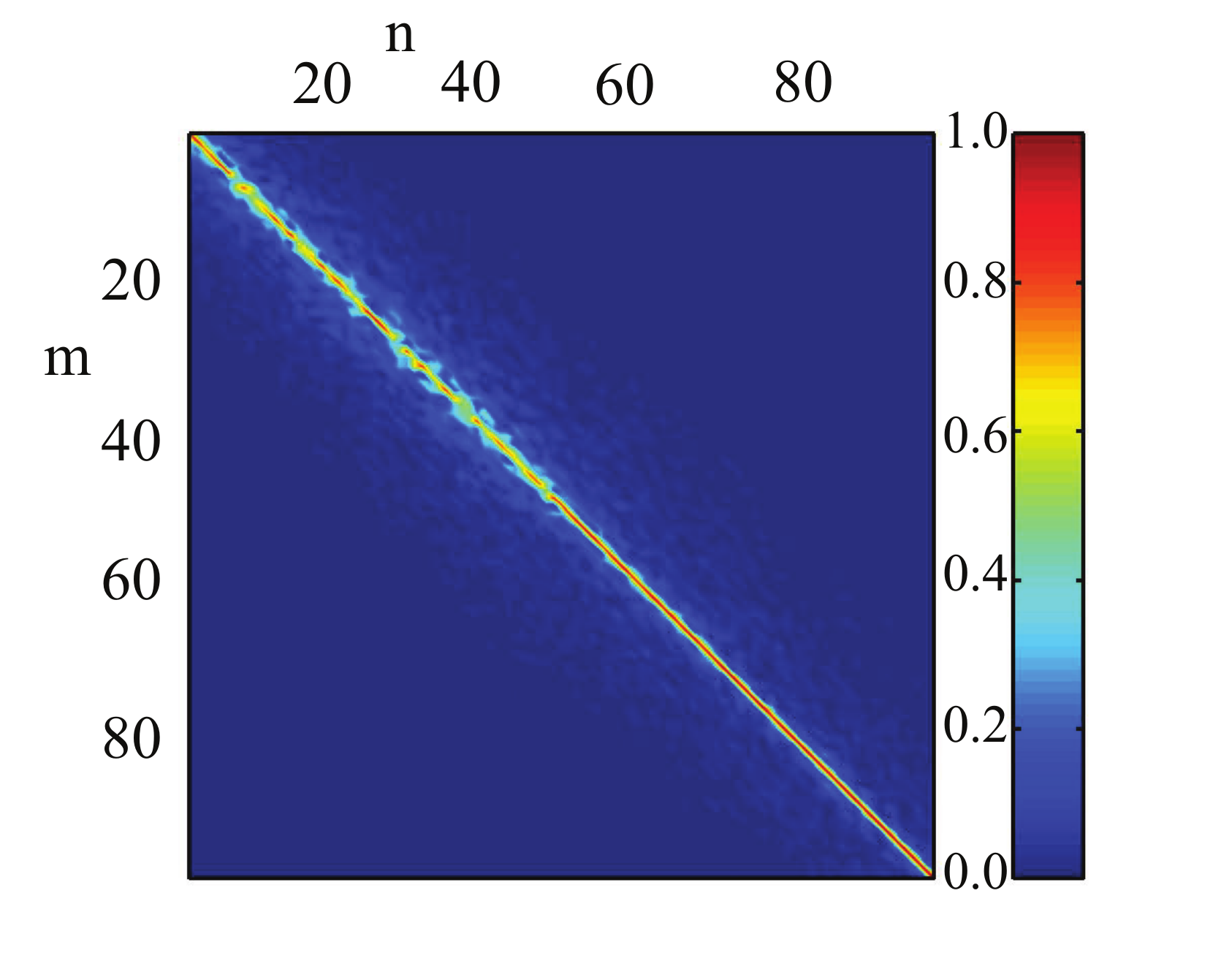}} 
     \subfigure{
          \includegraphics[width= 3.5 in,angle=0]{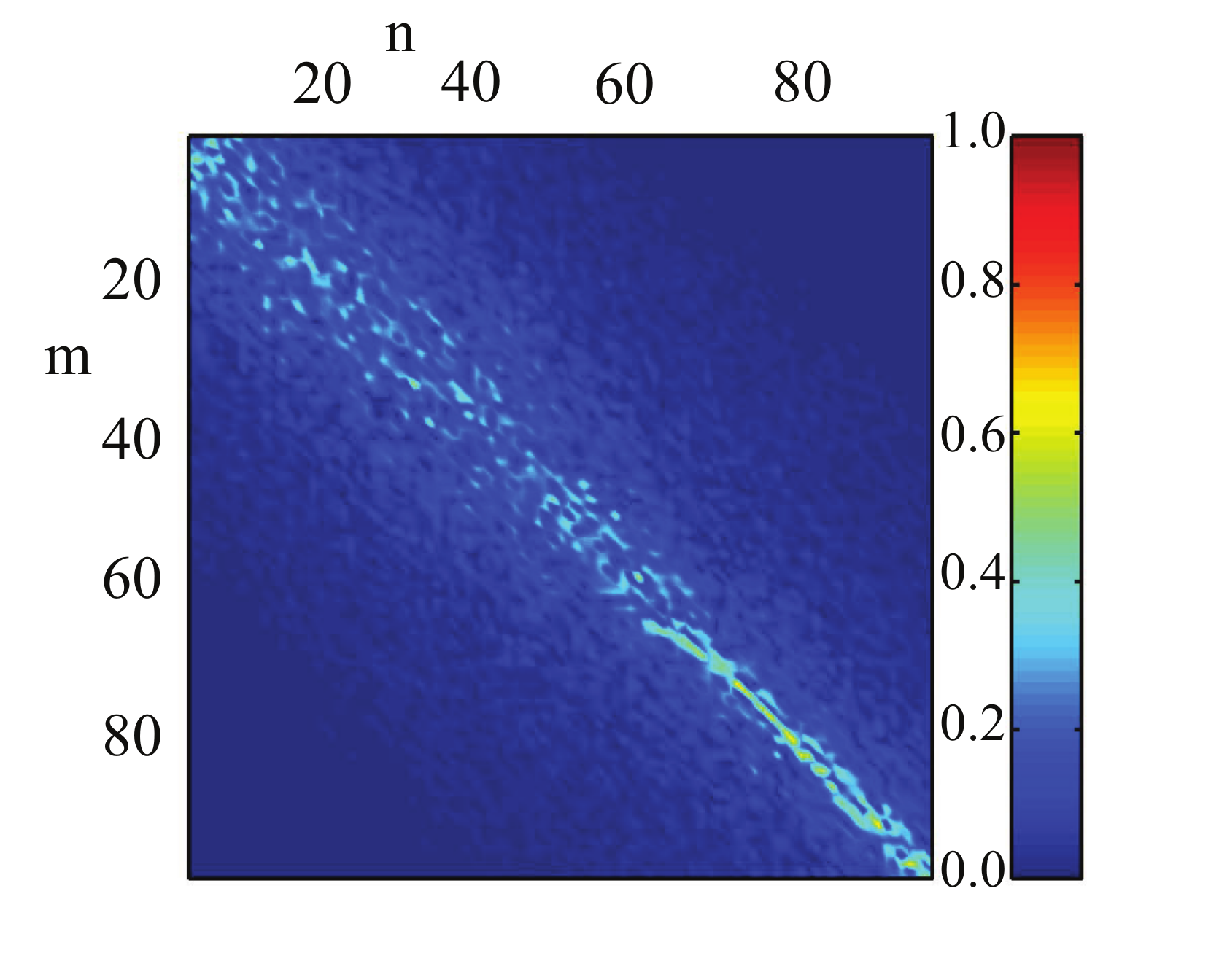}}
\caption{$U$ for propagation through a single internal wave field is illustrated as a color plot of the magnitudes of the matrix elements $|U_{mn}|$.  The top panel is for $50$ km propagation, and the bottom panel for $1000$ km.}
\label{fig2}
\end{figure}

One might anticipate that the mode-mixing from scattering cannot involve modes separated too greatly in mode number rendering the matrix banded, at least initially in the propagation.  Figure~\ref{fig2} illustrates  
the bandedness of $U$ due to the internal wave field.  Thus, the variance of the $A_{mn}$ decreases rapidly away from the diagonal.  The matrix possesses a rough translational invariance along the diagonal and the matrix element 
\begin{figure}[ht]
\includegraphics[width=3.2 in]{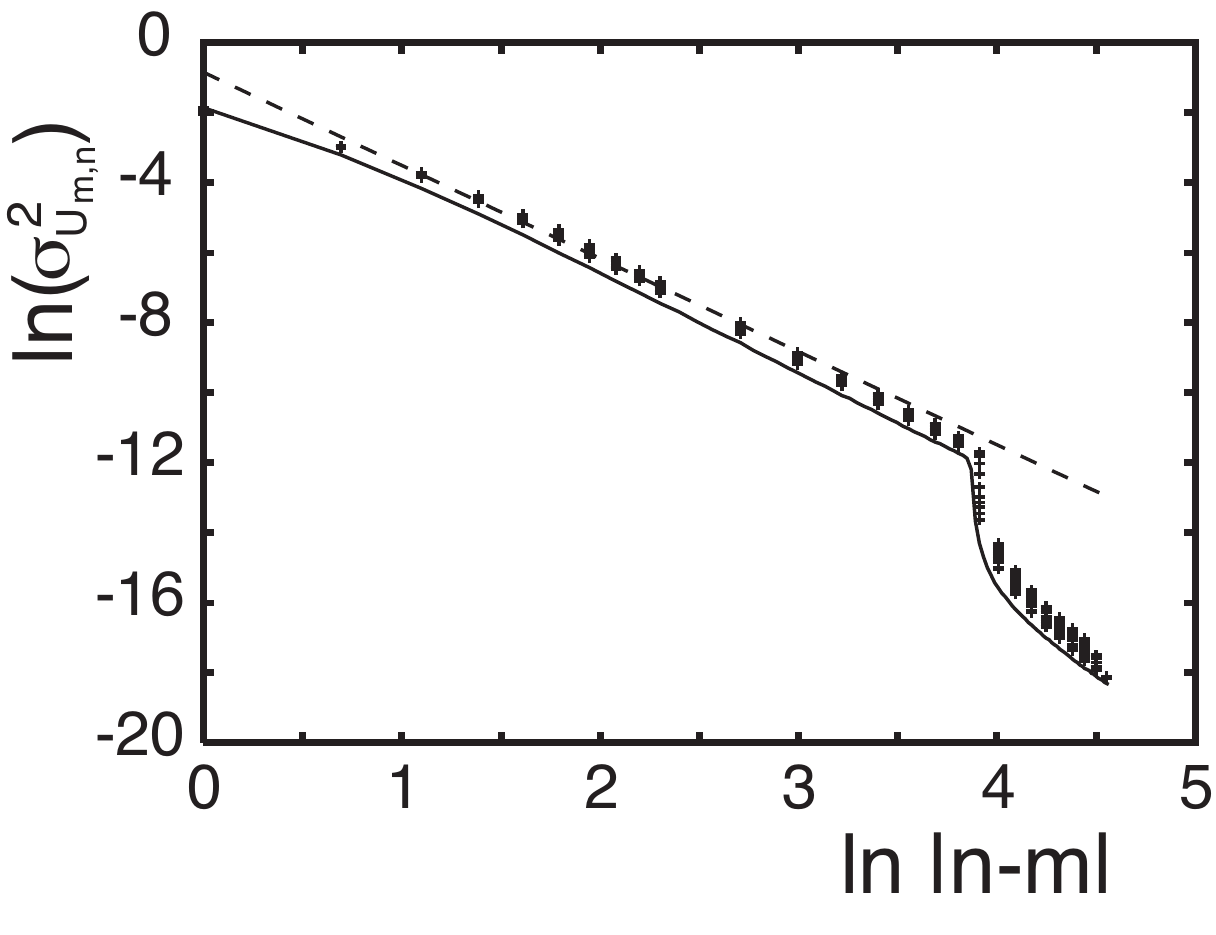}
\caption{The variance of the off-diagonal matrix elements of $U$ as a function of distance from the matrix diagonal.  The solid line is the result of a perturbation theory not explicitly discussed in the text.  The dashed line indicates the power-law decay exponent.}
\label{fig3}
\end{figure}
variance is mostly a function of $|n-m|$.  Figure~\ref{fig3} displays the behavior of the band, which has a power law decrease of exponent $\approx 1.3$ for the standard deviation up to a shoulder near $|n-m|\approx 50$, where it suddenly drops off much more sharply.  Further investigation of the shoulder is needed to ensure that it is not an artifact of the internal waves construction method.

The ensemble of Eq.~(\ref{ensemble}) with the variances of Fig.~\ref{fig3} has three features distinguishing it from the power-law random banded matrices introduced for investigations of the Anderson metal-insulator transition~\cite{Mirlin96}.  First, the ensemble is unitary; see Ref.~\cite{Bandyopadhyay10} for introduction and discussion of a unitary ensemble.  Second the diagonal elements retain a deterministic function and importance depending on the value of $\epsilon$.  Finally, there is a shoulder in the band width.  The band width exponent is roughly independent of wave vector and just a bit greater than unity, which makes this ensemble most similar to those possessing localization and super-diffusive wave packet spreading at short times~\cite{Mirlin96}.  Thus, the ensemble is consistent with localization in mode number, which allows for the well known possibility of isolating and measuring early arrival structures~\cite{Worcester99,Colosi99}.  Were the internal waves or other scattering mechanisms in the ocean of a different character with a slower decay away from the diagonal, that would not have been the case.  The bandwidth decay exponent is little changed by multiplication of $U_b$, at sufficiently long propagation range the width will grow sufficiently to make wave energy hit the ocean's surface and bottom, which will then strip out energy.  

\begin{figure}
\centering
     \subfigure{
          \includegraphics[width=3.5in,angle=0]{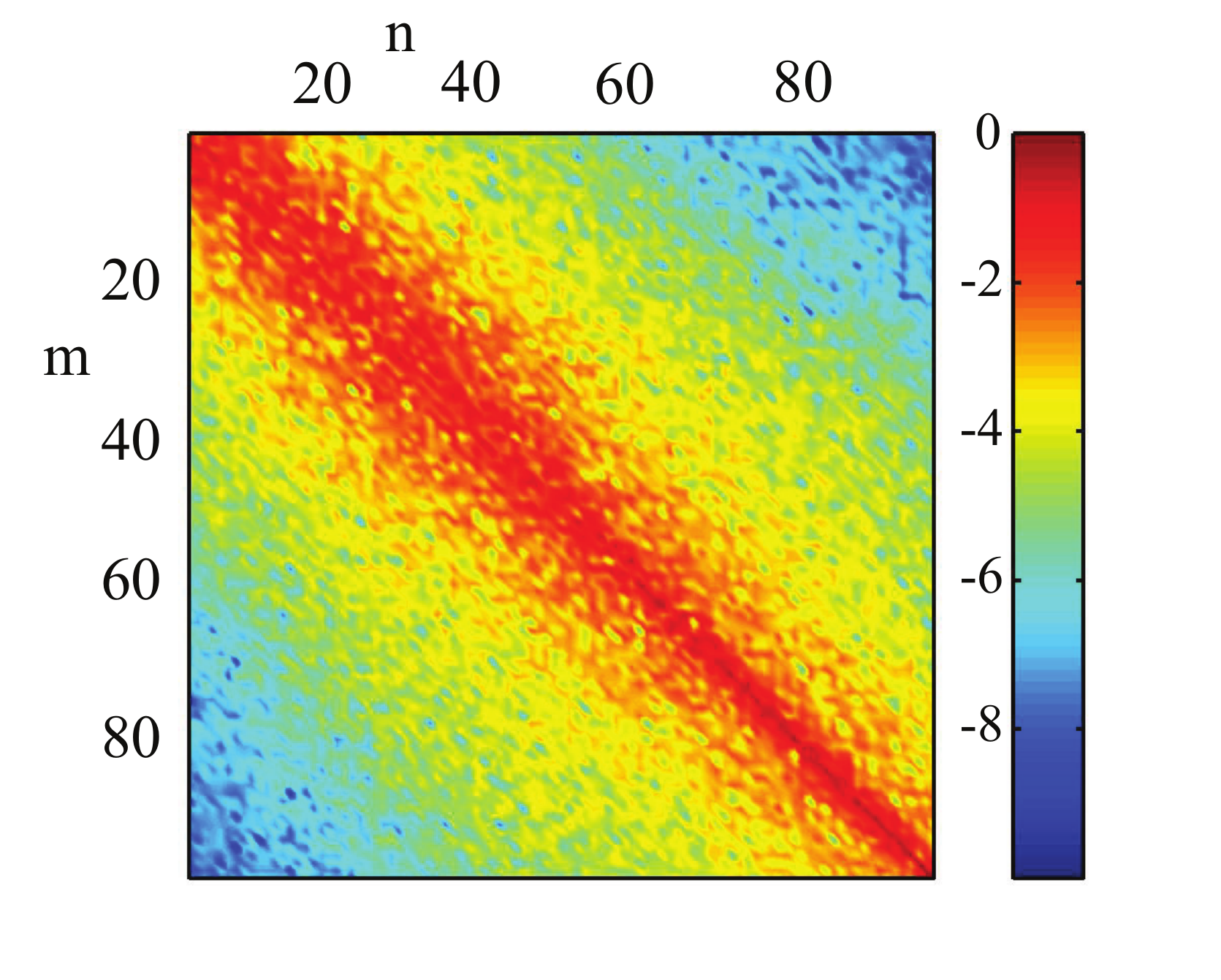}}\\
     \subfigure{
          \includegraphics[width=3.5in,angle=0]{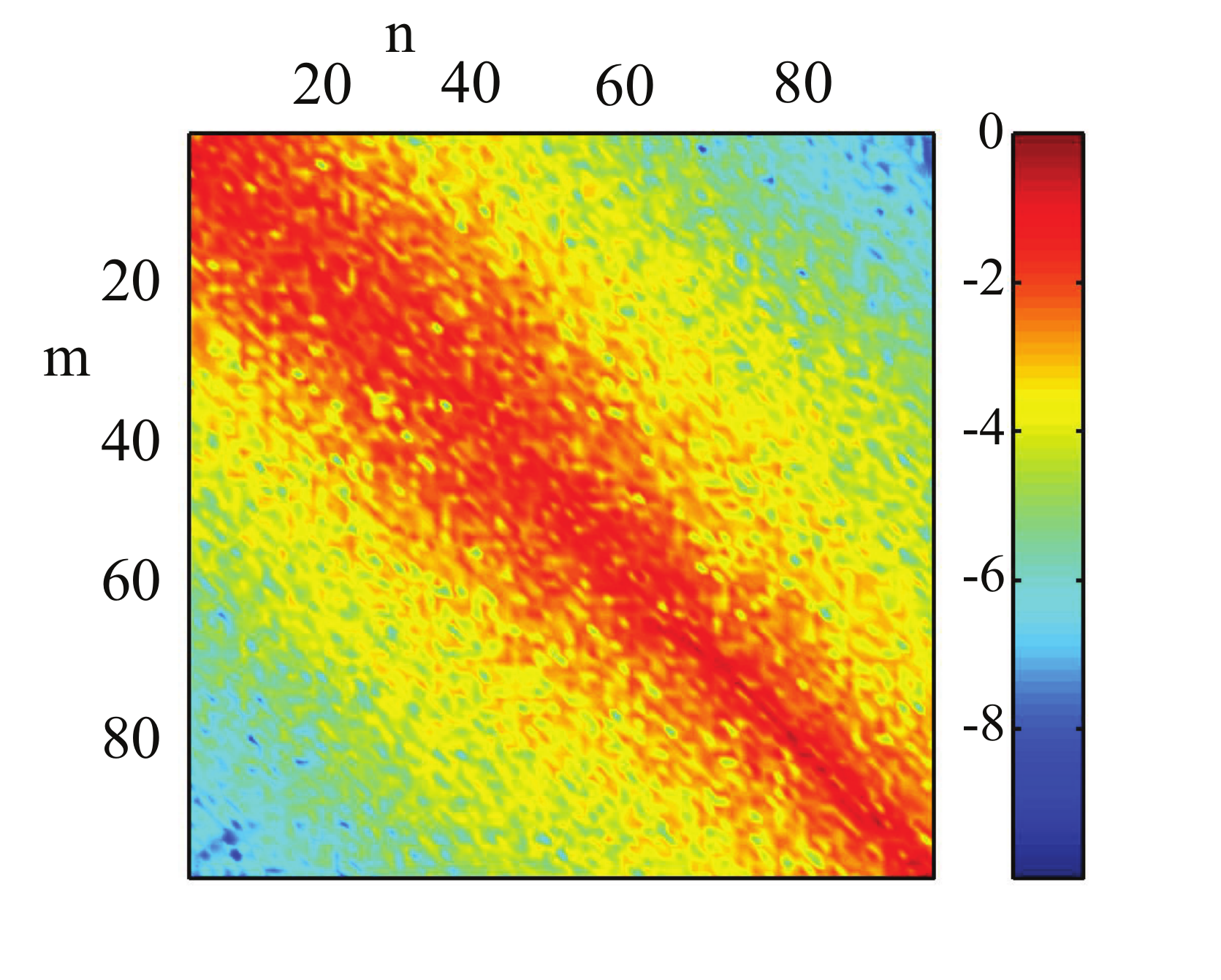}}
     \caption{Illustration of $U$ as a color plot of the logarithm of $|U_{mn}|$.  The upper panel results from a single ensemble realization, and the lower panel results from parabolic equation propagation through a single internal wave field.
  The propagation range is $r=1000$ km.}
\label{fig4}
\end{figure}

It is a somewhat brash assumption that dynamical correlations of neighboring matrix elements $U_{b,nm}$ can be completely ignored.  Nevertheless, it is seen in Fig.~\ref{fig4} that a typical unitary banded random ensemble matrix $U$ obtained as a product over building 
blocks and a $U$ constructed by propagation through a single realization of an internal wave field using the parabolic equation give the same general appearance, including band width, when propagated out to $1000$ km.  From this figure at least, there is no indication that this correlation information survives intact in propagation to long ranges.  Further details are to be published~\cite{Hegewisch11}.

In summary, for the first time a random matrix ensemble has been introduced for ocean acoustic propagation.  The ensemble is constructed for a building block unitary propagator,  and further propagation is by matrix multiplication of independently chosen $U_b$.  The matrix is banded  in mode number with an approximate power law whose exponent firmly places the ensemble in the class of those that exhibit localization and super-diffusion.  Unlike the ensembles introduced to study the Anderson metal-insulator transition~\cite{Mirlin96}, the ensemble here is unitary, has deterministic structure for the diagonal elements, and there is a cutoff/shoulder in the band width where the matrix elements fall off precipitously.  The relatively slow decay of a power law cutoff has ramifications for convergence of calculations in determining where one can make numerical approximations such as matrix truncations.  The ensemble construction is much faster than full wave propagation and for statistical purposes can be used to simulate the dynamics without having to explicitly construct the potential due to internal waves.  There have been a number of investigations of power-law banded random matrices in multiple contexts, and there exists some analytic analysis~\cite{Evers08} that may inform ocean acoustics applications.  It would be interesting to consider shallower water and derive ensembles with absorption and surface/bottom scattering.

\begin{acknowledgments}
We thank P.~A.~Mello and M.~G.~Brown for helpful discussions. Financial support from  the U.S.~National Science Foundation (PHY-0855337) is gratefully acknowledged.   Computational resources were supported in part by the U.S.~National Science Foundation through TeraGrid resources provided by NCSA.
\end{acknowledgments}

\bibliography{oceanacoustics,quantumchaos,rmtmodify,nano}

\begin{thebibliography}{26}%
\makeatletter
\providecommand \@ifxundefined [1]{%
 \@ifx{#1\undefined}
}%
\providecommand \@ifnum [1]{%
 \ifnum #1\expandafter \@firstoftwo
 \else \expandafter \@secondoftwo
 \fi
}%
\providecommand \@ifx [1]{%
 \ifx #1\expandafter \@firstoftwo
 \else \expandafter \@secondoftwo
 \fi
}%
\providecommand \natexlab [1]{#1}%
\providecommand \enquote  [1]{``#1''}%
\providecommand \bibnamefont  [1]{#1}%
\providecommand \bibfnamefont [1]{#1}%
\providecommand \citenamefont [1]{#1}%
\providecommand \href@noop [0]{\@secondoftwo}%
\providecommand \href [0]{\begingroup \@sanitize@url \@href}%
\providecommand \@href[1]{\@@startlink{#1}\@@href}%
\providecommand \@@href[1]{\endgroup#1\@@endlink}%
\providecommand \@sanitize@url [0]{\catcode `\\12\catcode `\$12\catcode
  `\&12\catcode `\#12\catcode `\^12\catcode `\_12\catcode `\%12\relax}%
\providecommand \@@startlink[1]{}%
\providecommand \@@endlink[0]{}%
\providecommand \url  [0]{\begingroup\@sanitize@url \@url }%
\providecommand \@url [1]{\endgroup\@href {#1}{\urlprefix }}%
\providecommand \urlprefix  [0]{URL }%
\providecommand \Eprint [0]{\href }%
\@ifxundefined \urlstyle {%
  \providecommand \doi  [0]{\begingroup \@sanitize@url \@doi}%
  \providecommand \@doi [1]{\endgroup \@@startlink {\doibase
  #1}doi:\discretionary {}{}{}#1\@@endlink }%
}{%
  \providecommand \doi  [0]{doi:\discretionary{}{}{}\begingroup
  \urlstyle{rm}\Url }%
}%
\providecommand \doibase [0]{http://dx.doi.org/}%
\providecommand \Doi [0]{\begingroup \@sanitize@url \@Doi }%
\providecommand \@Doi  [1]{\endgroup\@@startlink{\doibase#1}\@@Doi}%
\providecommand \@@Doi [1]{#1\@@endlink}%
\providecommand \selectlanguage [0]{\@gobble}%
\providecommand \bibinfo  [0]{\@secondoftwo}%
\providecommand \bibfield  [0]{\@secondoftwo}%
\providecommand \translation [1]{[#1]}%
\providecommand \BibitemOpen [0]{}%
\providecommand \bibitemStop [0]{}%
\providecommand \bibitemNoStop [0]{.\EOS\space}%
\providecommand \EOS [0]{\spacefactor3000\relax}%
\providecommand \BibitemShut  [1]{\csname bibitem#1\endcsname}%
\bibitem [{\citenamefont {Flatt\'e}\ \emph {et~al.}(1979)\citenamefont
  {Flatt\'e}, \citenamefont {Dashen}, \citenamefont {Munk},\ and\ \citenamefont
  {Zachariasen}}]{Flatte79}%
  \BibitemOpen
  \bibfield  {author} {\bibinfo {author} {\bibfnamefont {S.~M.}\ \bibnamefont
  {Flatt\'e}}, \bibinfo {author} {\bibfnamefont {R.}~\bibnamefont {Dashen}},
  \bibinfo {author} {\bibfnamefont {W.~H.}\ \bibnamefont {Munk}}, \ and\
  \bibinfo {author} {\bibfnamefont {F.}~\bibnamefont {Zachariasen}},\
  }\href@noop {} {\emph {\bibinfo {title} {Sound Transmission through a
  Fluctuating Ocean}}}\ (\bibinfo  {publisher} {Cambridge University Press},\
  \bibinfo {address} {Cambridge},\ \bibinfo {year} {1979})\BibitemShut
  {NoStop}%
\bibitem [{\citenamefont {Munk}\ \emph {et~al.}(1995)\citenamefont {Munk},
  \citenamefont {Worcester},\ and\ \citenamefont {Wuncsh}}]{Munk95}%
  \BibitemOpen
  \bibfield  {author} {\bibinfo {author} {\bibfnamefont {W.~H.}\ \bibnamefont
  {Munk}}, \bibinfo {author} {\bibfnamefont {P.}~\bibnamefont {Worcester}}, \
  and\ \bibinfo {author} {\bibfnamefont {C.}~\bibnamefont {Wuncsh}},\
  }\href@noop {} {\emph {\bibinfo {title} {Ocean Acoustic Tomography}}}\
  (\bibinfo  {publisher} {Cambridge University Press},\ \bibinfo {address}
  {Cambridge},\ \bibinfo {year} {1995})\BibitemShut {NoStop}%
\bibitem [{\citenamefont {Palmer}\ \emph {et~al.}(1988)\citenamefont {Palmer},
  \citenamefont {Brown}, \citenamefont {Tappert},\ and\ \citenamefont
  {Bezdek}}]{Palmer88}%
  \BibitemOpen
  \bibfield  {author} {\bibinfo {author} {\bibfnamefont {D.~R.}\ \bibnamefont
  {Palmer}}, \bibinfo {author} {\bibfnamefont {M.~G.}\ \bibnamefont {Brown}},
  \bibinfo {author} {\bibfnamefont {F.~D.}\ \bibnamefont {Tappert}}, \ and\
  \bibinfo {author} {\bibfnamefont {H.~F.}\ \bibnamefont {Bezdek}},\
  }\href@noop {} {\bibfield  {journal} {\bibinfo  {journal}
  {Geophys.~Res.~Lett.},\ }\textbf {\bibinfo {volume} {15}},\ \bibinfo {pages}
  {569} (\bibinfo {year} {1988})}\BibitemShut {NoStop}%
\bibitem [{\citenamefont {Palmer}\ \emph {et~al.}(1991)\citenamefont {Palmer},
  \citenamefont {Georges},\ and\ \citenamefont {Jones}}]{Palmer91}%
  \BibitemOpen
  \bibfield  {author} {\bibinfo {author} {\bibfnamefont {D.~R.}\ \bibnamefont
  {Palmer}}, \bibinfo {author} {\bibfnamefont {T.~M.}\ \bibnamefont {Georges}},
  \ and\ \bibinfo {author} {\bibfnamefont {R.~M.}\ \bibnamefont {Jones}},\
  }\href@noop {} {\bibfield  {journal} {\bibinfo  {journal}
  {Comput.~Phys.~Commun.},\ }\textbf {\bibinfo {volume} {65}},\ \bibinfo
  {pages} {219} (\bibinfo {year} {1991})}\BibitemShut {NoStop}%
\bibitem [{\citenamefont {Smith}\ \emph
  {et~al.}(1992){\natexlab{a}}\citenamefont {Smith}, \citenamefont {Brown},\
  and\ \citenamefont {Tappert}}]{Smith92a}%
  \BibitemOpen
  \bibfield  {author} {\bibinfo {author} {\bibfnamefont {K.~B.}\ \bibnamefont
  {Smith}}, \bibinfo {author} {\bibfnamefont {M.~G.}\ \bibnamefont {Brown}}, \
  and\ \bibinfo {author} {\bibfnamefont {F.~D.}\ \bibnamefont {Tappert}},\
  }\href@noop {} {\bibfield  {journal} {\bibinfo  {journal}
  {J.~Acoust.~Soc.~Am.},\ }\textbf {\bibinfo {volume} {91}},\ \bibinfo {pages}
  {1939} (\bibinfo {year} {1992}{\natexlab{a}})}\BibitemShut {NoStop}%
\bibitem [{\citenamefont {Smith}\ \emph
  {et~al.}(1992){\natexlab{b}}\citenamefont {Smith}, \citenamefont {Brown},\
  and\ \citenamefont {Tappert}}]{Smith92b}%
  \BibitemOpen
  \bibfield  {author} {\bibinfo {author} {\bibfnamefont {K.~B.}\ \bibnamefont
  {Smith}}, \bibinfo {author} {\bibfnamefont {M.~G.}\ \bibnamefont {Brown}}, \
  and\ \bibinfo {author} {\bibfnamefont {F.~D.}\ \bibnamefont {Tappert}},\
  }\href@noop {} {\bibfield  {journal} {\bibinfo  {journal}
  {J.~Acoust.~Soc.~Am.},\ }\textbf {\bibinfo {volume} {91}},\ \bibinfo {pages}
  {1950} (\bibinfo {year} {1992}{\natexlab{b}})}\BibitemShut {NoStop}%
\bibitem [{\citenamefont {Brown}\ \emph {et~al.}(2003)\citenamefont {Brown},
  \citenamefont {Colosi}, \citenamefont {Tomsovic}, \citenamefont
  {Virovlyansky}, \citenamefont {Wolfson},\ and\ \citenamefont
  {Zaslavsky}}]{Brown03}%
  \BibitemOpen
  \bibfield  {author} {\bibinfo {author} {\bibfnamefont {M.~G.}\ \bibnamefont
  {Brown}}, \bibinfo {author} {\bibfnamefont {J.~A.}\ \bibnamefont {Colosi}},
  \bibinfo {author} {\bibfnamefont {S.}~\bibnamefont {Tomsovic}}, \bibinfo
  {author} {\bibfnamefont {A.~L.}\ \bibnamefont {Virovlyansky}}, \bibinfo
  {author} {\bibfnamefont {M.~A.}\ \bibnamefont {Wolfson}}, \ and\ \bibinfo
  {author} {\bibfnamefont {G.~M.}\ \bibnamefont {Zaslavsky}},\ }\href@noop {}
  {\bibfield  {journal} {\bibinfo  {journal} {J.~Acoust.~Soc.~Am.},\ }\textbf
  {\bibinfo {volume} {113}},\ \bibinfo {pages} {2533} (\bibinfo {year}
  {2003})},\ \bibinfo {note} {nlin.CD/0109027}\BibitemShut {NoStop}%
\bibitem [{\citenamefont {Beron-Vera}\ \emph {et~al.}(2003)\citenamefont
  {Beron-Vera}, \citenamefont {Brown}, \citenamefont {Colosi}, \citenamefont
  {Tomsovic}, \citenamefont {Virovlyansky}, \citenamefont {Wolfson},\ and\
  \citenamefont {Zaslavsky}}]{Beronvera03}%
  \BibitemOpen
  \bibfield  {author} {\bibinfo {author} {\bibfnamefont {F.~J.}\ \bibnamefont
  {Beron-Vera}}, \bibinfo {author} {\bibfnamefont {M.~G.}\ \bibnamefont
  {Brown}}, \bibinfo {author} {\bibfnamefont {J.~A.}\ \bibnamefont {Colosi}},
  \bibinfo {author} {\bibfnamefont {S.}~\bibnamefont {Tomsovic}}, \bibinfo
  {author} {\bibfnamefont {A.~L.}\ \bibnamefont {Virovlyansky}}, \bibinfo
  {author} {\bibfnamefont {M.~A.}\ \bibnamefont {Wolfson}}, \ and\ \bibinfo
  {author} {\bibfnamefont {G.~M.}\ \bibnamefont {Zaslavsky}},\ }\href@noop {}
  {\bibfield  {journal} {\bibinfo  {journal} {J.~Acoust.~Soc.~Am.},\ }\textbf
  {\bibinfo {volume} {114}},\ \bibinfo {pages} {1226} (\bibinfo {year}
  {2003})},\ \bibinfo {note} {arXiv:0301026 [nlin.CD]}\BibitemShut {NoStop}%
\bibitem [{\citenamefont {Tomsovic}\ and\ \citenamefont
  {Brown}(2010)}]{Tomsovic10}%
  \BibitemOpen
  \bibfield  {author} {\bibinfo {author} {\bibfnamefont {S.}~\bibnamefont
  {Tomsovic}}\ and\ \bibinfo {author} {\bibfnamefont {M.~G.}\ \bibnamefont
  {Brown}},\ }in\ \href@noop {} {\emph {\bibinfo {booktitle} {New directions in
  linear acoustics and vibration: random matrix theory, quantum chaos and
  complexity}}},\ \bibinfo {editor} {edited by\ \bibinfo {editor}
  {\bibfnamefont {R.}~\bibnamefont {Weaver}}\ and\ \bibinfo {editor}
  {\bibfnamefont {M.}~\bibnamefont {Wright}}}\ (\bibinfo  {publisher}
  {Cambridge University Press},\ \bibinfo {address} {New York},\ \bibinfo
  {year} {2010})\ pp.\ \bibinfo {pages} {169--187}\BibitemShut {NoStop}%
\bibitem [{\citenamefont {Brody}\ \emph {et~al.}(1981)\citenamefont {Brody},
  \citenamefont {Flores}, \citenamefont {French}, \citenamefont {Mello},
  \citenamefont {Pandey},\ and\ \citenamefont {Wong}}]{Brody81}%
  \BibitemOpen
  \bibfield  {author} {\bibinfo {author} {\bibfnamefont {T.~A.}\ \bibnamefont
  {Brody}}, \bibinfo {author} {\bibfnamefont {J.}~\bibnamefont {Flores}},
  \bibinfo {author} {\bibfnamefont {J.~B.}\ \bibnamefont {French}}, \bibinfo
  {author} {\bibfnamefont {P.~A.}\ \bibnamefont {Mello}}, \bibinfo {author}
  {\bibfnamefont {A.}~\bibnamefont {Pandey}}, \ and\ \bibinfo {author}
  {\bibfnamefont {S.~S.~M.}\ \bibnamefont {Wong}},\ }\href@noop {} {\bibfield
  {journal} {\bibinfo  {journal} {Rev.~Mod.~Phys.},\ }\textbf {\bibinfo
  {volume} {53}},\ \bibinfo {pages} {385} (\bibinfo {year} {1981})}\BibitemShut
  {NoStop}%
\bibitem [{\citenamefont {Bohigas}\ \emph {et~al.}(1984)\citenamefont
  {Bohigas}, \citenamefont {Giannoni},\ and\ \citenamefont
  {Schmit}}]{Bohigas84}%
  \BibitemOpen
  \bibfield  {author} {\bibinfo {author} {\bibfnamefont {O.}~\bibnamefont
  {Bohigas}}, \bibinfo {author} {\bibfnamefont {M.-J.}\ \bibnamefont
  {Giannoni}}, \ and\ \bibinfo {author} {\bibfnamefont {C.}~\bibnamefont
  {Schmit}},\ }\href@noop {} {\bibfield  {journal} {\bibinfo  {journal}
  {Phys.~Rev.~Lett.},\ }\textbf {\bibinfo {volume} {52}},\ \bibinfo {pages} {1}
  (\bibinfo {year} {1984})}\BibitemShut {NoStop}%
\bibitem [{\citenamefont {Mehta}(2004)}]{Mehta04}%
  \BibitemOpen
  \bibfield  {author} {\bibinfo {author} {\bibfnamefont {M.~L.}\ \bibnamefont
  {Mehta}},\ }\href@noop {} {\emph {\bibinfo {title} {Random Matrices (Third
  Edition)}}}\ (\bibinfo  {publisher} {Elsevier},\ \bibinfo {address}
  {Amsterdam},\ \bibinfo {year} {2004})\BibitemShut {NoStop}%
\bibitem [{\citenamefont {Weaver}(1989)}]{Weaver89}%
  \BibitemOpen
  \bibfield  {author} {\bibinfo {author} {\bibfnamefont {R.~L.}\ \bibnamefont
  {Weaver}},\ }\href@noop {} {\bibfield  {journal} {\bibinfo  {journal}
  {J.~Acoust.~Soc.~Am.},\ }\textbf {\bibinfo {volume} {85}},\ \bibinfo {pages}
  {1005} (\bibinfo {year} {1989})}\BibitemShut {NoStop}%
\bibitem [{\citenamefont {Wright}\ and\ \citenamefont
  {Weaver}(2010)}]{Wright10}%
  \BibitemOpen
  \bibinfo {editor} {\bibfnamefont {M.}~\bibnamefont {Wright}}\ and\ \bibinfo
  {editor} {\bibfnamefont {R.}~\bibnamefont {Weaver}},\ eds.,\ \href@noop {}
  {\emph {\bibinfo {title} {New directions in linear acoustics and vibration:
  quantum chaos, random matrix theory and complexity}}}\ (\bibinfo  {publisher}
  {Cambridge University Press},\ \bibinfo {address} {Cambridge},\ \bibinfo
  {year} {2010})\BibitemShut {NoStop}%
\bibitem [{\citenamefont {Munk}(1974)}]{Munk74}%
  \BibitemOpen
  \bibfield  {author} {\bibinfo {author} {\bibfnamefont {W.~H.}\ \bibnamefont
  {Munk}},\ }\href@noop {} {\bibfield  {journal} {\bibinfo  {journal}
  {J.~Acoust.~Soc.~Am.},\ }\textbf {\bibinfo {volume} {55}},\ \bibinfo {pages}
  {220} (\bibinfo {year} {1974})}\BibitemShut {NoStop}%
\bibitem [{\citenamefont {Garrett}\ and\ \citenamefont
  {Munk}(1979)}]{Garrett79}%
  \BibitemOpen
  \bibfield  {author} {\bibinfo {author} {\bibfnamefont {C.~J.~R.}\
  \bibnamefont {Garrett}}\ and\ \bibinfo {author} {\bibfnamefont {W.~H.}\
  \bibnamefont {Munk}},\ }\href@noop {} {\bibfield  {journal} {\bibinfo
  {journal} {Annu.~Rev.~Fluid Mech.},\ }\textbf {\bibinfo {volume} {11}},\
  \bibinfo {pages} {339} (\bibinfo {year} {1979})}\BibitemShut {NoStop}%
\bibitem [{\citenamefont {Colosi}\ and\ \citenamefont
  {Brown}(1998)}]{Colosi98}%
  \BibitemOpen
  \bibfield  {author} {\bibinfo {author} {\bibfnamefont {J.~A.}\ \bibnamefont
  {Colosi}}\ and\ \bibinfo {author} {\bibfnamefont {M.~G.}\ \bibnamefont
  {Brown}},\ }\href@noop {} {\bibfield  {journal} {\bibinfo  {journal}
  {J.~Acoust.~Soc.~Am.},\ }\textbf {\bibinfo {volume} {103}},\ \bibinfo {pages}
  {2232} (\bibinfo {year} {1998})}\BibitemShut {NoStop}%
\bibitem [{\citenamefont {Dozier}\ and\ \citenamefont
  {Tappert}(1978){\natexlab{a}}}]{Dozier78}%
  \BibitemOpen
  \bibfield  {author} {\bibinfo {author} {\bibfnamefont {L.~B.}\ \bibnamefont
  {Dozier}}\ and\ \bibinfo {author} {\bibfnamefont {F.~D.}\ \bibnamefont
  {Tappert}},\ }\href@noop {} {\bibfield  {journal} {\bibinfo  {journal}
  {J.~Acoust.~Soc.~Am.},\ }\textbf {\bibinfo {volume} {63}},\ \bibinfo {pages}
  {353} (\bibinfo {year} {1978}{\natexlab{a}})}\BibitemShut {NoStop}%
\bibitem [{\citenamefont {Dozier}\ and\ \citenamefont
  {Tappert}(1978){\natexlab{b}}}]{Dozier78b}%
  \BibitemOpen
  \bibfield  {author} {\bibinfo {author} {\bibfnamefont {L.~B.}\ \bibnamefont
  {Dozier}}\ and\ \bibinfo {author} {\bibfnamefont {F.~D.}\ \bibnamefont
  {Tappert}},\ }\href@noop {} {\bibfield  {journal} {\bibinfo  {journal}
  {J.~Acoust.~Soc.~Am.},\ }\textbf {\bibinfo {volume} {63}},\ \bibinfo {pages}
  {533} (\bibinfo {year} {1978}{\natexlab{b}})}\BibitemShut {NoStop}%
\bibitem [{\citenamefont {Froufe-Perez}\ \emph {et~al.}(2007)\citenamefont
  {Froufe-Perez}, \citenamefont {Yepez}, \citenamefont {Mello},\ and\
  \citenamefont {Saenz}}]{Perez07}%
  \BibitemOpen
  \bibfield  {author} {\bibinfo {author} {\bibfnamefont {L.~S.}\ \bibnamefont
  {Froufe-Perez}}, \bibinfo {author} {\bibfnamefont {M.}~\bibnamefont {Yepez}},
  \bibinfo {author} {\bibfnamefont {P.~A.}\ \bibnamefont {Mello}}, \ and\
  \bibinfo {author} {\bibfnamefont {J.~J.}\ \bibnamefont {Saenz}},\ }\href@noop
  {} {\bibfield  {journal} {\bibinfo  {journal} {Phys.~Rev.~E},\ }\textbf
  {\bibinfo {volume} {75}},\ \bibinfo {pages} {031113} (\bibinfo {year}
  {2007})}\BibitemShut {NoStop}%
\bibitem [{\citenamefont {Mirlin}\ \emph {et~al.}(1996)\citenamefont {Mirlin},
  \citenamefont {Fyodorov}, \citenamefont {Dittes}, \citenamefont {Quezada},\
  and\ \citenamefont {Seligman}}]{Mirlin96}%
  \BibitemOpen
  \bibfield  {author} {\bibinfo {author} {\bibfnamefont {A.~D.}\ \bibnamefont
  {Mirlin}}, \bibinfo {author} {\bibfnamefont {Y.~V.}\ \bibnamefont
  {Fyodorov}}, \bibinfo {author} {\bibfnamefont {F.~M.}\ \bibnamefont
  {Dittes}}, \bibinfo {author} {\bibfnamefont {J.}~\bibnamefont {Quezada}}, \
  and\ \bibinfo {author} {\bibfnamefont {T.~H.}\ \bibnamefont {Seligman}},\
  }\href@noop {} {\bibfield  {journal} {\bibinfo  {journal} {Phys.~Rev.~E},\
  }\textbf {\bibinfo {volume} {54}},\ \bibinfo {pages} {3221} (\bibinfo {year}
  {1996})}\BibitemShut {NoStop}%
\bibitem [{\citenamefont {Bandyopadhyay}\ \emph {et~al.}(2010)\citenamefont
  {Bandyopadhyay}, \citenamefont {Wang},\ and\ \citenamefont
  {Gong}}]{Bandyopadhyay10}%
  \BibitemOpen
  \bibfield  {author} {\bibinfo {author} {\bibfnamefont {J.~N.}\ \bibnamefont
  {Bandyopadhyay}}, \bibinfo {author} {\bibfnamefont {J.}~\bibnamefont {Wang}},
  \ and\ \bibinfo {author} {\bibfnamefont {J.}~\bibnamefont {Gong}},\
  }\href@noop {} {\bibfield  {journal} {\bibinfo  {journal} {Phys.~Rev.~E},\
  }\textbf {\bibinfo {volume} {81}},\ \bibinfo {pages} {066212} (\bibinfo
  {year} {2010})}\BibitemShut {NoStop}%
\bibitem [{\citenamefont {Worcester}\ \emph {et~al.}(1999)\citenamefont
  {Worcester}, \citenamefont {Cornuelle}, \citenamefont {Dzieciuch},
  \citenamefont {Munk}, \citenamefont {Howe}, \citenamefont {A.Mercer},
  \citenamefont {Spindel}, \citenamefont {Colosi}, \citenamefont {Metzger},
  \citenamefont {Birdsall},\ and\ \citenamefont {Baggeroer}}]{Worcester99}%
  \BibitemOpen
  \bibfield  {author} {\bibinfo {author} {\bibfnamefont {P.~F.}\ \bibnamefont
  {Worcester}}, \bibinfo {author} {\bibfnamefont {B.~D.}\ \bibnamefont
  {Cornuelle}}, \bibinfo {author} {\bibfnamefont {M.~A.}\ \bibnamefont
  {Dzieciuch}}, \bibinfo {author} {\bibfnamefont {W.~H.}\ \bibnamefont {Munk}},
  \bibinfo {author} {\bibfnamefont {B.~M.}\ \bibnamefont {Howe}}, \bibinfo
  {author} {\bibfnamefont {J.}~\bibnamefont {A.Mercer}}, \bibinfo {author}
  {\bibfnamefont {R.~C.}\ \bibnamefont {Spindel}}, \bibinfo {author}
  {\bibfnamefont {J.~A.}\ \bibnamefont {Colosi}}, \bibinfo {author}
  {\bibfnamefont {K.}~\bibnamefont {Metzger}}, \bibinfo {author} {\bibfnamefont
  {T.}~\bibnamefont {Birdsall}}, \ and\ \bibinfo {author} {\bibfnamefont
  {A.~B.}\ \bibnamefont {Baggeroer}},\ }\href@noop {} {\bibfield  {journal}
  {\bibinfo  {journal} {J.~Acoust.~Soc.~Am.},\ }\textbf {\bibinfo {volume}
  {105}},\ \bibinfo {pages} {3185} (\bibinfo {year} {1999})}\BibitemShut
  {NoStop}%
\bibitem [{\citenamefont {Colosi}\ \emph {et~al.}(1999)\citenamefont {Colosi},
  \citenamefont {Scheer}, \citenamefont {Flatt\'{e}}, \citenamefont
  {Cornuelle}, \citenamefont {Dzieciuch}, \citenamefont {Munk}, \citenamefont
  {Worcester}, \citenamefont {Howe}, \citenamefont {A.Mercer}, \citenamefont
  {Spindel}, \citenamefont {Metzger}, \citenamefont {Birdsall},\ and\
  \citenamefont {Baggeroer}}]{Colosi99}%
  \BibitemOpen
  \bibfield  {author} {\bibinfo {author} {\bibfnamefont {J.~A.}\ \bibnamefont
  {Colosi}}, \bibinfo {author} {\bibfnamefont {E.~K.}\ \bibnamefont {Scheer}},
  \bibinfo {author} {\bibfnamefont {S.~M.}\ \bibnamefont {Flatt\'{e}}},
  \bibinfo {author} {\bibfnamefont {B.~D.}\ \bibnamefont {Cornuelle}}, \bibinfo
  {author} {\bibfnamefont {M.~A.}\ \bibnamefont {Dzieciuch}}, \bibinfo {author}
  {\bibfnamefont {W.~H.}\ \bibnamefont {Munk}}, \bibinfo {author}
  {\bibfnamefont {P.~F.}\ \bibnamefont {Worcester}}, \bibinfo {author}
  {\bibfnamefont {B.~M.}\ \bibnamefont {Howe}}, \bibinfo {author}
  {\bibfnamefont {J.}~\bibnamefont {A.Mercer}}, \bibinfo {author}
  {\bibfnamefont {R.~C.}\ \bibnamefont {Spindel}}, \bibinfo {author}
  {\bibfnamefont {K.}~\bibnamefont {Metzger}}, \bibinfo {author} {\bibfnamefont
  {T.}~\bibnamefont {Birdsall}}, \ and\ \bibinfo {author} {\bibfnamefont
  {A.~B.}\ \bibnamefont {Baggeroer}},\ }\href@noop {} {\bibfield  {journal}
  {\bibinfo  {journal} {J.~Acoust.~Soc.~Am.},\ }\textbf {\bibinfo {volume}
  {105}},\ \bibinfo {pages} {3202} (\bibinfo {year} {1999})}\BibitemShut
  {NoStop}%
\bibitem [{\citenamefont {Hegewisch}\ and\ \citenamefont
  {Tomsovic}(2011)}]{Hegewisch11}%
  \BibitemOpen
  \bibfield  {author} {\bibinfo {author} {\bibfnamefont {K.~C.}\ \bibnamefont
  {Hegewisch}}\ and\ \bibinfo {author} {\bibfnamefont {S.}~\bibnamefont
  {Tomsovic}},\ }\href@noop {} {\bibfield  {journal} {\bibinfo  {journal}
  {Phys.~Rev.~E}} (\bibinfo {year} {2011})},\ \bibinfo {note} {to be
  submitted}\BibitemShut {NoStop}%
\bibitem [{\citenamefont {Evers}\ and\ \citenamefont {Mirlin}(2008)}]{Evers08}%
  \BibitemOpen
  \bibfield  {author} {\bibinfo {author} {\bibfnamefont {F.}~\bibnamefont
  {Evers}}\ and\ \bibinfo {author} {\bibfnamefont {A.~D.}\ \bibnamefont
  {Mirlin}},\ }\href@noop {} {\bibfield  {journal} {\bibinfo  {journal}
  {Rev.~Mod.~Phys.},\ }\textbf {\bibinfo {volume} {80}},\ \bibinfo {pages}
  {1355} (\bibinfo {year} {2008})}\BibitemShut {NoStop}%
\end{thebibliography}%

\end{document}